\documentclass{article}

\usepackage[final]{neurips_2024}
\usepackage{graphicx}

\usepackage[utf8]{inputenc} % allow utf-8 input
\usepackage[T1]{fontenc}    % use 8-bit T1 fonts
\usepackage{hyperref}       % hyperlinks
\usepackage{url}            % simple URL typesetting
\usepackage{booktabs}       % professional-quality tables
\usepackage{amsfonts}       % blackboard math symbols
\usepackage{nicefrac}       % compact symbols for 1/2, etc.
\usepackage{microtype}      % microtypography
\usepackage{xcolor}         % colors

\title{Revival: Collaborative Artistic Creation through Human-AI Interactions in Musical Creativity}

\newcommand*\samethanks[1][\value{footnote}]{\footnotemark[#1]}
\author{
  Keon Ju M. Lee\thanks{Metacreation Lab for Creative AI: \url{https://www.metacreation.net/}} \\
  School of Interactive Arts and Technology\\
  Simon Fraser University\\
  Vancouver, British Columbia, Canada \\
  \texttt{keon\_maverick@sfu.ca} \\
  \And
  Philippe Pasquier\samethanks \\
  School of Interactive Arts and Technology \\
  Simon Fraser University \\
  Vancouver, British Columbia, Canada \\
  \texttt{pasquier@sfu.ca} \\
  \And
  Jun Yuri \\
  Independent \\
  Vancouver, British Columbia, Canada \\
  \texttt{mail@amagi.dev} \\
}

\begin{document}

\maketitle

\begin{abstract}
\textbf{Revival} is an innovative live audiovisual performance and music improvisation by our artist collective K-Phi-A, blending human and AI musicianship to create electronic music with audio-reactive visuals. The performance features real-time co-creative improvisation between a percussionist, an electronic music artist, and AI musical agents. Trained in works by deceased composers and the collective’s compositions, these agents dynamically respond to human input and emulate complex musical styles. An AI-driven visual synthesizer, guided by a human VJ, produces visuals that evolve with the musical landscape. \textbf{Revival} showcases the potential of AI and human collaboration in improvisational artistic creation.
\end{abstract}

\section{Introduction}
\textbf{Artwork concept:}
Revival is an artwork and innovative live audiovisual performance by the artist collective K-Phi-A, which integrates human and machine musicianship through artificial intelligence to produce electronic music and its audio-reactive visuals. The performance is centred on real-time co-creative improvisation between human performers—a percussionist and an electronic music performer—and AI musical agents, such as MASOM [1] and SpireMuse [2]. These musical agents are trained on the works of deceased composers and compositions created by the collective, enabling them to respond dynamically to human input and emulate complex musical styles. An AI-driven visual synthesizer trained on public domain images, Autolume [3], generates visuals responsive to the evolving musical landscape with a human VJ. This performance exemplifies the potential of AI in co-creative artistic creation, demonstrating how AI and human artists can collaboratively create in a live music improvisation context.

%Central to Revival is the principle of real-time co-creative improvisation between AI agents and human artists, as opposed to offline computer-assisted composition. The musical agents are trained on small personalized datasets, including those of deceased electroacoustic composers. Crafting models and exploring affordances for the agents are crucial aspects of collaborative art creation with AI agents, enabling adaptation to specific musical and artistic contexts. These musical agents dynamically respond to human performances, replicating the stylistic intricacies of deceased composers through musique concrète. Simultaneously, the Autolume agent generates immersive audiovisuals reactive to musical expressions trained on images in the public domain.

\textbf{Artist collective:}
The K-Phi-A consists of percussionist Keon Ju Maverick Lee, who collaborates with musical agents; electronic music performer Philippe Pasquier, who utilizes a DIY software sampler named Kenaxis and agents; and VJ Amagi, who operates the Autolume audiovisual synthesizer.

\textbf{Dr. Philippe Pasquier} is a scientific researcher, a multidisciplinary media artist, an educator, and a community builder. He is the director of the Metacreation Lab for Creative AI and a Professor at Simon Fraser University (SFU), School for Interactive Arts and Technology in Vancouver, Canada. He is pursuing a multidisciplinary research-creation program focused on generative systems and applied AI for creative tasks there. His music AI algorithms have reached millions through collaborations with the software industry (e.g., Steinberg, Ableton, Teenage Engineering, Elias, etc.), and his artworks have been shown on six continents.

\textbf{Keon Ju Maverick Lee} is an electronic percussionist, drummer, music composer, and sound artist focusing on research-creation for designing a musical agent and interactive music systems to accompany drum performances. He has been active as a drummer in the local music scene in British Columbia, Canada. Keon is a Ph.D. student in the Metacreation Lab for Creative AI at Simon Fraser University. He is a lab instructor for the Sound Design course at SFU and computer music courses at the University of Victoria. As an academic, he publishes research-creation papers in music and AI. He collaborates with colleagues at the Centre for Digital Music in London, UK, and IRCAM in Paris, France. Recently, he has been collaborating with a media artist to get involved in the ReVerie project performed at Ars Electronica and SIGGRAPH [4] in 2024.

\textbf{VJ Amagi} is a VJ (Video Jockey) / graphics programmer based in Vancouver, Canada. He performs with real-time visuals and creates art installations using TouchDesigner and Unity. Additionally, he develops VJ systems for shader live coding performances. At MUTEK.jp 2018, he led a workshop on visual effects using TouchDesigner.

The detailed performance setup can be found in Appendix A, and the showcase of the artwork Revival is accessible on the Metacreation Lab website\footnote{\url{https://www.metacreation.net/projects/revival-art}}.

\section{Human-Machine Interaction for Music Improvisation with Audiovisual}
The core element of our human-machine interaction framework is the machine listening module [5] embedded within the AI musical agent system [6]. This module integrates with our conductor software environment, utilizing Chataigne, AbletonLive, and the AI-driven visual synthesizer Autolume. The artistic exploration examines the possibilities of musical practices facilitated by a co-creative framework [11], employing a research-creation methodology [12] to investigate the dynamic interplay between music and reactive audiovisuals as a medium. This inquiry operates at the nexus of music improvisation, audiovisual art, and media art, emphasizing human-AI collaboration in co-creation. Our approach advocates for a small data mindset [10], highlighting the responsible and ethical utilization of data resources. Small data offers ethical advantages by enhancing transparency around copyrighted material for artists and musicians. Focusing on curated datasets limits excessive data scraping and promotes responsible resource use, aligning with principles of data minimalism and artist protection. Furthermore, adopting a small data mindset is more environmentally sustainable, demanding significantly fewer computational resources for training processes than big data systems.

\textbf{Offline machine listening:} Musical agents engage with a percussionist and electronic music performances within our interactive music system, leveraging custom machine listening algorithms. This system employs both low-level audio features and high-level affective features to enhance the interactivity of the musical agents. Each audio segment is annotated with a feature vector consisting of 55 dimensions, which includes duration, mean and standard deviation of loudness, Mel-Frequency Cepstral Coefficients (MFCC), fundamental frequency, chroma, and high-level affective features such as valence and arousal [1]. To achieve more precise melodic and harmonic analysis, a large FFT (Fast Fourier Transform [7]) window size of 8192 samples combined with a smaller hop size of 512 samples was employed, providing high-frequency resolution. This configuration was chosen based on experimental results. We incorporate chroma features, which represent pitch histograms corresponding to the 12 chromatic scale notes, significantly enhancing the musical agents' ability to detect harmonic and melodic characteristics. Chroma features enhance melodic classification and reduce pitch errors in audio segments with multiple notes, providing a discrete representation that facilitates the identification of harmonic content. This approach also improves clustering within the self-organizing map [8], even when the calculated pitch inaccurately reflects tonality. In our affective computing module of the agent, Valence and arousal, as defined by the circumplex model of affect [9], are computed from a multivariate linear regression with coefficients of low-level audio features and further enhance the system's capacity to recognize the emotional characteristics of sound samples.

\textbf{Real-time machine listening:} During runtime, the machine listening algorithm continuously segments the user's input stream into segments aligned with those in the reference corpus. The listening module can adjust feature weights, subsequently influencing the matching algorithms. In SpireMuse, four influence dimensions—rhythmic, spectral, melodic, and harmonic—are available as parameters [2]. The rhythmic parameter prioritizes the duration feature, while the spectral parameter emphasizes Mel-Frequency Cepstral Coefficients (MFCC) features. The melodic parameter focuses on the fundamental frequency, and the harmonic parameter highlights chroma features. MASOM is designed to handle primary musical passages within the musical agent system, whereas SpireMuse accompanies a percussionist and electronic music performer, leveraging its real-time machine listening module.

We also explored the creative potential of our musical agent system, MACAT, through its integration into our artwork. Building upon the MASOM agent, MACAT enhances real-time concatenative sound synthesis and introduces advanced visualization capabilities. While MASOM emphasizes experimental music using self-organizing maps and Variable Markov Models, MACAT incorporates concatenative synthesis and the Factor Oracle for superior pattern recognition and generative outputs. These innovations make MACAT highly adaptable and responsive in live performances, offering expanded timbral flexibility and enabling dynamic creative possibilities.

\textbf{Audiovisual:} In the interactive component of our audiovisual system, spectral and Bark coefficients are extracted from the audio outputs of Keon, Philippe, and the musical agents. These coefficients are transmitted via Open Sound Control (OSC) messages to Autolume and VJ Amagi, enabling the generation of reactive visuals in real-time, which are shaped by VJ Amagi’s artistic techniques for visual music. Furthermore, the extracted audio features are conveyed to the Digital Multiplex (DMX) interface to control reactive lighting, facilitating visual representation for each performer. The reactive lighting system enhances the visual interpretation of each performer's actions by colour-mapping specific audio features to the lighting system, thus enriching the audience’s performance experience. 

\textbf{Conductor:} We utilize Chataigne software to configure our conductor system, which orchestrates communication, sound/visual parameter control, and timelines for our structured live music improvisation and audiovisual performance. The performance comprises multiple musical and visual themes, each employing distinct sound parameters for musical agents, percussion samples, audio samples, and visual elements. The conductor coordinates these elements in real-time through OSC messages to interact with the musical agents, visual synthesizer, and DMX lighting systems.  

\section{Research-Creation Methodology in Artwork}
The integration of research-creation methodology in music improvisation and performances [12] involving co-creation with musical AI agents and human artists is increasingly recognized as a vital area of exploration within contemporary music studies. This methodology emphasizes the collaborative and iterative nature of artistic practice, allowing for a deeper understanding of both the creative processes involved and the implications of AI in music.

Research-creation methodology facilitates a nuanced exploration of the interplay between human musicians and AI agents, particularly in live performance contexts. For instance, studies have shown that AI can act as a reactive and creative partner in musical improvisation, enhancing the performance experience by introducing novel elements that challenge traditional notions of authorship and creativity [13,14]. This co-creative dynamic not only enriches the musical output but also prompts musicians to rethink their roles as creators in the face of intelligent systems that can generate music in real-time [15]. The collaborative nature of this process is essential, as it allows for the blending of human intuition with AI's computational capabilities, leading to innovative musical expressions that neither could achieve independently [16,17].

Moreover, the ethical dimensions of integrating AI into music creation cannot be overlooked. Engaging musicians and audiences in the research process helps to address potential conflicts and ensures that the technology serves the artistic community effectively [18,19]. By incorporating systematic user studies and feedback mechanisms, researchers can better understand the emotional and aesthetic responses elicited by AI-generated music, which is crucial for refining these technologies to meet the needs of human artists [20,21]. This participatory approach aligns with the principles of research-creation, emphasizing the importance of collaboration and exploration in the artistic process.

The implications of AI in music extend beyond mere technical innovation; they also challenge existing paradigms of creativity and authorship. The emergence of AI-driven music generation systems has led to discussions about the perceived autonomy of these systems and their impact on human creativity [16,22]. As AI continues to evolve, it raises questions about the nature of musical composition and the role of the composer, prompting a re-evaluation of what it means to create music in a digital age [23,24]. The research-creation methodology provides a framework for investigating these complex issues, allowing for an analysis of the artistic, ethical, and social dimensions of AI in music.

In conclusion, the importance of research-creation methodology in music improvisation and performances involving AI agents lies in its capacity to foster collaboration, address ethical concerns, and challenge traditional notions of creativity. By engaging in this iterative process, musicians and researchers can explore the rich possibilities that arise from the intersection of human artistry and artificial intelligence, ultimately leading to a deeper understanding of both fields.

\section{Challenges}
Collaborative audiovisual performances involving musical AI agents and visual synthesizers with human artists face significant challenges in achieving seamless integration between audio and visual modalities. Synchronization of timing, dynamics, and artistic intent across these mediums is crucial but technically demanding, particularly in real-time contexts. Any latency or misalignment can disrupt the artistic flow and compromise the audience’s experience. The complexity of managing multiple systems, such as Max/MSP for music and TouchDesigner for visuals, further heightens the technical demands on performers and developers.

Balancing creative autonomy between AI agents and human artists adds another layer of complexity. While AI needs to generate innovative and meaningful contributions, it must align with the human artist’s vision to ensure artistic cohesion. This balance is especially challenging during improvisational performances, where AI’s unpredictability can necessitate quick adaptation from human performers. Careful design of interaction paradigms is essential to prevent the performance from feeling disjointed, ensuring that the audio-visual relationship remains fluid and engaging.

Finally, audience perception plays a critical role in the success of such performances. Effectively communicating the role of AI while emphasizing the collaborative nature of the performance can enhance audience understanding and appreciation. Artists must navigate these technical, creative, and perceptual challenges to create compelling, cohesive audiovisual experiences that highlight the potential of human-AI collaboration in the arts.
\section{Conclusion}
The collaborative artistic creation explored in the "Revival" performance demonstrates the profound potential of integrating human and AI interactions in musical creativity. Through a research-creation methodology, the project highlights the importance of dynamic co-creation, where AI agents and human performers engage in a seamless exchange of ideas and creativity. The innovative use of AI musical agents, such as MASOM and SpireMuse, alongside the Autolume visual synthesizer, emphasizes how AI can act as a reactive and creative partner, enriching live performances with novel elements that challenge traditional notions of authorship and creativity.

This work underscores the role of human agency in shaping the outcomes of human-AI collaboration while leveraging AI’s computational capabilities for innovative musical and visual expressions. The integration of curated datasets and ethical approaches, like a "small data" mindset, ensures that these artistic explorations are conducted responsibly and sustainably. The performance not only redefines the boundaries of improvisational music but also provokes deeper reflections on the evolving relationship between technology and artistry.

Ultimately, the Revival project showcases the transformative potential of AI in the arts, offering a glimpse into a future where human-AI collaboration becomes a standard practice for creative exploration. By addressing technical challenges, fostering interdisciplinary collaboration, and ensuring ethical practices, this work paves the way for a deeper understanding of the possibilities and implications of AI in the cultural and creative domains.
\section*{References}

{
\small

[1] Tatar, K. \& Pasquier, P. (2017). MASOM: A Musical Agent Architecture based on Self-Organizing Maps, Affective Computing, and Variable Markov Models. In the proceedings of the Conference on International Workshop on Musical Metacreation (MuMe).

[2] Thelle, N. J. W., \& Pasquier, P. (2021). Spire Muse: A Virtual Musical Partner for Creative Brainstorming. In the proceedings of the Conference on New Interfaces for Musical Expression (NIME).

[3] Kraasch, J., \& Pasquier, P. (2022). Autolume-Live: Turning GANs into a Live VJing tool. In the proceedings of the Conference on Computation, Communication, Aesthetics \& X (xCoAx).

[4] Liu, P., \& Lee, K. J. (2024). ReVerie. In ACM SIGGRAPH 2024 Immersive Pavilion.

[5] Rowe, R. (1992). Interactive music systems: machine listening and composing. MIT press.

[6] Tatar, K., \& Pasquier, P. (2019). Musical agents: A typology and state of the art towards musical metacreation. Journal of New Music Research.

[7] Nussbaumer, H. J., \& Nussbaumer, H. J. (1982). The fast Fourier transform (pp. 80-111). Springer Berlin Heidelberg.

[8] Kohonen, T. (1990). The self-organizing map. In the proceedings of the conference on the Institute of Electrical and Electronics Engineers (IEEE).

[9] Russell, J. A. (1980). A circumplex model of affect. Journal of personality and social psychology

[10] Vigliensoni, G., Perry, P., \& Fiebrink, R. (2022). A small-data mindset for generative AI creative work. In the proceedings of the Machine Learning for Creativity and Design Workshop at the conference on Neural Information Processing Systems (NIPS).

[11] Thelle, N. J. W., \& Wærstad, B. I. G. (2023). Co-Creatives Spaces: The machine as a collaborator. In Proceedings of the Conference on New Interfaces for Musical Expression (NIME).

[12] Stévance, S., \& Lacasse, S. (2017). creation in Music and the Arts: Towards a Collaborative Interdiscipline. Routledge.

[13] Vear, C. (2021). Creative AI and musicking robots. Frontiers in Robotics and AI, 8, 631752.

[14] Marley, W., \& Ward, N. (2015, June). Tightly coupled agents in live performance metacreations. In Proceedings of the 2015 ACM SIGCHI Conference on Creativity and Cognition (pp. 299-302).

[15] Ben-Tal, O., Harris, M. T., \& Sturm, B. L. (2021). How music AI is useful: engagements with composers, performers and audiences. Leonardo, 54(5), 510-516.

[16] Verma, S. (2021). Artificial intelligence and music: History and the future perceptive. International Journal of Applied Research, 7(2), 272-275.

[17] Smith, J., \& Freeman, J. (2021, October). Effects of deep neural networks on the perceived creative autonomy of a generative musical system. In Proceedings of the AAAI Conference on Artificial Intelligence and Interactive Digital Entertainment (Vol. 17, No. 1, pp. 91-98).

[18] Deruty, E., Grachten, M., Lattner, S., Nistal, J., \& Aouameur, C. (2022). On the development and practice of AI technology for contemporary popular music production. Transactions of the International Society for Music Information Retrieval, 5(1), 35-50.

[19] Holzapfel, A., Sturm, B., \& Coeckelbergh, M. (2018). Ethical dimensions of music information retrieval technology. Transactions of the International Society for Music Information Retrieval, 1(1), 44-55.

[20] Sturm, B. L., Iglesias, M., Ben-Tal, O., Miron, M., \& Gómez, E. (2019, September). Artificial intelligence and music: open questions of copyright law and engineering praxis. In Arts (Vol. 8, No. 3, p. 115). MDPI.

[21] Dash, A., \& Agres, K. (2024). AI-based affective music generation systems: A review of methods and challenges. ACM Computing Surveys, 56(11), 1-34.

[22] Miranda, E. R. (Ed.). (2021). Handbook of artificial intelligence for music. Cham: Springer.

[23] Shank, D. B., Stefanik, C., Stuhlsatz, C., Kacirek, K., \& Belfi, A. M. (2023). AI composer bias: Listeners like music less when they think it was composed by an AI. Journal of Experimental Psychology: Applied, 29(3), 676.

[24] Carnovalini, F., \& Rodà, A. (2020). Computational creativity and music generation systems: An introduction to the state of the art. Frontiers in Artificial Intelligence, 3, 14.

%%%%%%%%%%%%%%%%%%%%%%%%%%%%%%%%%%%%%%%%%%%%%%%%%%%%%%%%%%%%
\clearpage
\appendix

\section{Detailed Setup for Performance}
We offer a comprehensive setup for our music improvisation and audiovisual performance, detailing the necessary components and configurations. Given the ease of organizing audiovisual performances in Vancouver, where all required equipment and resources are readily accessible, we are prepared to perform if needed.

\subsection{General}
\begin{itemize}
    \item \textbf{Title of piece}: Revival
    \item \textbf{Performer}: Electronic percussionist, live electronic performer, AI musical agents, AI audiovisual agent and VJ
    \item \textbf{Duration}: The performance is designed to last approximately 30 minutes (flexible); however, we can accommodate a shorter demonstration of under 10 minutes if necessary. This shorter session would feature a simplified setup, particularly suitable for contexts where time is limited. We recognize that the NIPS workshop may not be a conventional venue for performing arts, and we are prepared to adjust the presentation accordingly.
    \item \textbf{Contact person}: Keon Ju Maverick Lee 
(Email: keon\_maverick@sfu.ca)
\end{itemize}

\subsection{Technical requirements with stage plot}
List of instruments/equipment/material provided by K-Phi-A: 
\begin{itemize}
    \item DIY instruments designed in Max/Msp as software
    \item Electronic multi-pad percussion with a pedal with AbletonLive including Max4Live
    \item Minimum 4 or 5 laptops would be brought by performers: 1 laptop for MASOM and SpireMuse musical agents and electronic percussion pads for Keon, 1 laptop for Philippe’s electronic instrument, 1 laptop for Autolume audio-visual agent, and 1 laptop for VJ Amagi.
\end{itemize}

Number of people on stage: 
3 humans + 3 AI agents (co-creation)

Desired stage plot / specific lighting and/or video needs: 
The artwork would be a 30-minute audiovisual performance combining live music and synchronized video projection. The adaptable musical structure allows for easy adjustments to suit audience engagement and event objectives, so the length of the performance is flexible.

\begin{figure}[!ht]
\centering
 \includegraphics[width=0.7\columnwidth]{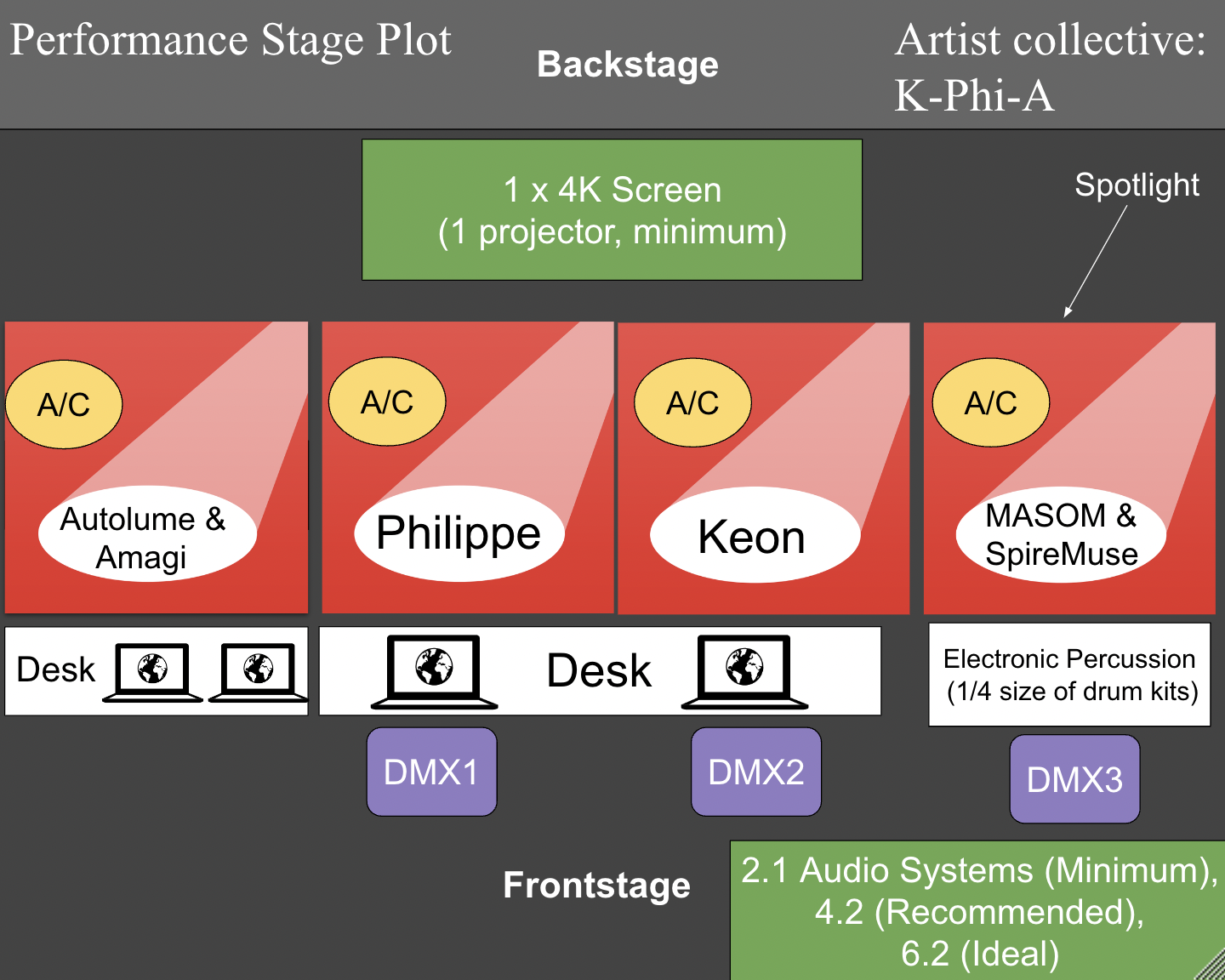}
 \caption{Stage plot for our performance setup.}
 \label{fig:stage-plot}
\end{figure}

\begin{table}[ht]
\begin{center}
\begin{tabular}{cccc}
    \toprule
    \textbf{Audio Channel} & \textbf{Input Source} & \textbf{Visual \& Light} & \textbf{Input Source} \\
    \midrule
    1 & Philippe & 4K Projector & VJ Amagi \\ 
    2 & Keon & DMX-1 & Philippe \\
    3 & MASOM & DMX-2 & MASOM \\
    4 & SpireMuse & DMX-3 & Keon \\
    \bottomrule
\end{tabular}
\end{center}
\caption{List of input sources for audio and visual, including lights for DMX.  \label{input}}
\end{table}
* Note: A/Cs are necessary to charge our 4 laptops and electronic multi-drum pads. 
Summary: our minimum setup is 2.1 audio systems with one 4K Projector screen.
Our optional setup is 4.2+ audio systems with two 4K projector screens and 3 DMX lights.

\textbf{Stage Plot Description (Figure \ref{fig:stage-plot}):} 
One (minimum) or two (optional) 4K HD projectors using HDMIs, 2.1 + audio system (minimum), and the projections are ideally positioned next to each other, ground to ceiling, behind the performers.
Equipment for 3 DMX channels (optional) for reactive lighting is needed. Each performer (Keon, Philippe, MASOM) will send DMX signals to our DMX interface, and the light will react to the amount of sound produced by each performer as part of performance visualization.
We will transmit our audio to the audio system for music and 1 HDMI output for interactive audiovisual by VJ Amagi to the technical team.
2.1 (Minimum), 2.2, 4.1, 4.2 (recommended), 6.1 or 6.2 audio systems (ideal) and ultra-sub bass frequencies are being used in this performance, so subwoofers are required. 
We will bring 4-5 laptops, so it would be ideal to have electric outlets to recharge to avoid potential power-off during the performance since laptops are the primary resources.
We are using our local router to communicate and interact with each other via Open Sound Control (OSC) messages.
Spotlight (optional): Keon, Philippe, MASOM \& SpireMuse, and Autolume with Amagi need a spotlight each.

There is no warning for audiovisual and music content.

For reference, we include the following information for software and hardware setup:

\textbf{Software:} The configuration utilizes various software tools, encompassing Max/MSP/Jitter, TouchDesigner, Chataigne, Ableton Live, Max for Live, and TouchOSC. Chataigne functions as the conductor software, managing communication within the interactive music system of musical agents. Max/MSP/Jitter is the foundational platform for implementing MASOM \& SpireMuse musical agents. Ableton Live and Max for Live are instrumental in designing the music system, integrating electronic drum and sampler instruments, including MIDI controllers, and configuring audio setups. TouchDesigner is employed for real-time audiovisual processing with the Autolume audiovisual agent and for developing the visual system. Additionally, TouchOSC enhances reactivity for interactive artwork in real-time with VJing, utilizing OSC and NDI signals transmitted through the local router for communication.

\textbf{Hardware:} The artistic and technical intricacies of our hardware setup and interactivity have been documented in a PDF format, encompassing details such as the technical rider and stage plot for our interactive audiovisual performance and artwork.

\end{document}